\documentclass[onecolumn]{emulateapj}
\usepackage{natbib}
\shortauthors{Chung, Lee, \& Koo}
\shorttitle{ Detection of planets in EWCP events}
\newcommand{\te}{t_{\rm E}}

\newcommand{\thetae}{\theta_{\rm E}}

\begin{document}
\title{Detection of planets in extremely weak central perturbation microlensing events via next-generation ground-based surveys}

\author{
Sun-Ju Chung\altaffilmark{1},
Chung-Uk Lee\altaffilmark{1},
and 
Jae-Rim Koo\altaffilmark{1}
 }
\altaffiltext{1}
{Korea Astronomy and Space Science Institute 776, Daedeokdae-ro,
Yuseong-Gu, Daejeon 305-348, Republic of Korea; sjchung@kasi.re.kr, leecu@kasi.re.kr, koojr@kasi.re.kr}


\begin{abstract}
Even though the recently discovered high-magnification event MOA-2010-BLG-311 had complete coverage over the peak, confident planet detection did not happen due to extremely weak central perturbations (fractional deviations of $\lesssim 2\%$).
For confident detection of planets in extremely weak central perturbation (EWCP) events, it is necessary to have both high cadence monitoring and high photometric accuracy better than those of current follow-up observation systems.
The next-generation ground-based observation project, KMTNet (Korea Microlensing Telescope Network), satisfies the conditions.
We estimate the probability of occurrence of EWCP events with fractional deviations of $\leq 2\%$ in high-magnification events and the efficiency of detecting planets in the EWCP events using the KMTNet.
From this study, we find that the EWCP events occur with a frequency of $> 50\%$ in the case of $\lesssim 100\ M_{\rm E}$ planets with separations of $0.2\ {\rm AU} \lesssim d \lesssim 20\ {\rm AU}$.
We find that for main-sequence and subgiant source stars, $\gtrsim 1\ M_{\rm E}$ planets in EWCP events with the deviations $\leq 2\%$ can be detected $> 50\%$ in a certain range that changes with the planet mass.
However, it is difficult to detect planets in EWCP events of bright stars like giant stars, because it is easy for KMTNet to be saturated around the peak of the events with a constant exposure time.
EWCP events are caused by close, intermediate, and wide planetary systems with low-mass planets and close and wide planetary systems with massive planets.
Therefore, we expect that a much greater variety of planetary systems than those already detected, which are mostly intermediate planetary systems regardless of the planet mass, will be significantly detected in the near future.
\end{abstract}

\section{INTRODUCTION}

High-magnification events for which the background source star passes close to the host star are very sensitive for detection of planets \citep{griest98}.
This is because the central caustic induced by a planet is formed near the host star and thus produces central perturbations near the peak of the lensing light curve.
\citet{rattenbury02} studied planet detectability in high-magnification events and they showed that Earth-mass planets may be detected with 1 m class telescopes.
Hence, current microlensing follow-up observations have focused on high-magnification events and  have resulted so far in the detection of 14 planets out of 25 extrasolar planets detected by microlensing (Udalski et al. 2005; Gould et al. 2006; Gaudi et al. 2008; Bennett et al. 2008; Dong et al. 2009; Janczak et al. 2010; Miyake et al. 2011; Bachelet et al. 2012; Yee et al. 2012; Han et al. 2013; Choi et al. 2013; Suzuki et al. 2013).

High-magnification events are sensitive to the diameter of the source star because the source star passes close to the central caustic.
If the source diameter is bigger than the central caustic and thus the finite-source effect is strong, central perturbations induced by the central caustic are greatly washed out, thus making it difficult to realize the existence of planets.
Events MOA-2007-BLG-400 (Subo et al. 2009), MOA-2008-BLG-310 (Janczak et al. 2010), and MOA-2010-BLG-311 (Yee et al. 2013) were high-magnification events with strong finite-source effects.
All three events had complete coverage over the peak, but a secure planet detection occurred in only two events, MOA-2007-BLG-400 and MOA-2008-BLG-310.
Even though the event MOA-2010-BLG-311 had complete coverage over the peak, central perturbations around the peak were extremely weak with a fractional deviation of $\lesssim 2\%$ so that it gave rise to a $\Delta \chi^2 \sim 80$ of the best-fit planetary lens model from the single lens model.
\citet{yee13} reported that the planetary signal of  $\Delta \chi^2 \sim 80$ is below the detection threshold range of $\Delta \chi^2 = 350-700$ suggested by \citet{gould10}, and thus it is difficult to claim a secure detection of the planet.
This suggests that extremely weak central perturbations (hereafter EWCPs) of the deviations $\lesssim 2\%$ produce planetary signals below the detection threshold and obstruct a confident detection of planets.
Current follow-up observations are intensively monitoring high-magnification events and their photometric error reaches $\sim 1\%$ at the peak, but it is not enough to get a confident detection of planets in EWCP events with deviations $\lesssim 2\%$, as shown in the event MOA-2010-BLG-311.
For confident planet detection in EWCP events, it is necessary to have both high cadence monitoring and high photometric accuracy better than those of current follow-up observation systems.
The next-generation ground-based observation project, KMTNet (Korea Microlensing Telescope Network), satisfies the conditions.
The KMTNet will use a 1.6 m wide-field telescope at each of three southern sites, Chile, South Africa, and Australia, to perform a 24 hr continuous observation toward the Galactic bulge \citep{kim10}.
Each telescope has a $18 {\rm K} \times 18 {\rm K}$ CCD camera that covers a field of view (FOV) of $2\arcdeg \times 2\arcdeg$ and it will observe four fields with a total FOV of $4\arcdeg \times 4\arcdeg$, in which each field will be monitored with an exposure time of about 2 minutes and a detector readout time of about 30 seconds giving a cadence of 10 minutes (Kim et al. 2010; Atwood et al. 2012).
Hence, KMTNet has high potential for the detection of planetary signals in EWCP events.
Here, we study how well planets in EWCP high-magnification events can be detected by the KMTNet.

This paper is organized as follows.
In \S\ 2, we briefly describe the properties of the central caustic induced by a planet.
In \S\ 3, we estimate the probability of occurrence of EWCP events with deviations $\leq 2\%$ in high-magnification events and the efficiency of detecting planets in the EWCP events using KMTNet.
In \S\ 4, we discuss the observational limitations and other potential studies of the KMTNet.
We summarize the results and conclude in \S\ 5.

\section{CENTRAL CAUSTIC}

In planetary lensing composed of a host star and a planet, the signal of the planet is a short-duration perturbation on the standard single lensing light curve of the host star.
The perturbation is caused by the central and planetary caustics, which are typically separated from each other.
The central caustic is always formed close to the host star and thus the perturbation by the central caustic occurs near the peak of the lensing light curve, while the planetary caustic is formed away from the host star and thus the perturbation by the planetary caustic can occur at any part of the light curve.

Central perturbations caused by the central caustic have generally a property of the $s \leftrightarrow 1/s$ degeneracy (Griest \& Safizadeh 1998; Dominik 1999).
The degeneracy arises due to the similarity in the size and shape of the central caustics for $s$ and $1/s$.
The duration of the central perturbations is proportional to the size of the central caustic.
The size of the central caustic defined by the separation of the cusps on the star-planet axis \citep{chung05} is expressed by
\begin{equation}
\Delta \xi \sim {4q \over {(s - 1/s)^2}}, 
\end{equation}
where $q$ is the planet/star mass ratio.
According to Equation (1), the size of the central caustic is $\propto s^2$ for $s \ll 1$ and is $\propto s^{-2}$ for $s \gg 1$.

The finite source effect for high-magnification events becomes important because the source star passes close to the central caustic, as mentioned before.
The magnification of a finite source corresponds to the magnification averaged over the source surface, i.e.,
\begin{equation}
A = {{\int^{\rho_\star}_{0} I(r)A_{p}(|\bold{r} -  \bold{r}_{L}|)rdr}\over{\int^{\rho_\star}_{0}I(r)rdr}},
\end{equation}
where $A_p$ is the point source magnification, $I(r)$ represents the source brightness profile, $\bold{r}$ is the vector to a position on the source star surface with respect to the center of the source star, $\bold{r}_L$ is the displacement vector of the source center with respect to the lens, and $\rho_\star$ is the source radius normalized to the Einstein radius of the lens system, $\thetae$, which is given by
\begin{equation}
\thetae = \sqrt{{ {4GM \over {c^2}} \left({1\over D_L} - {1\over D_S}\right)}},
\end{equation}
where $D_{\rm L}$ and $D_{\rm S}$ are the distances to the lens and the source from the observer, respectively.

\section{DETECTION EFFICIENCY}

\subsection{\it Probability}

Based on the result of the high-magnification event MOA-2010-BLG-311 (Yee et al. 2013) mentioned in Section 1, we choose that the threshold of EWCPs is a fractional deviation of $\delta = 2\%$, which is defined as
\begin{equation}
\delta = {A - A_{0} \over {A_0}}\ ,
\end{equation}
where $A$ and $A_0$ are the lensing magnifications with and without a planet, respectively.
To investigate the frequency of EWCP events in high-magnification events of $A_{\rm max} \geq 100$, we estimate the probability of occurrence of EWCP events with $\delta \leq 2\%$.
In consideration of typical Galactic bulge events, we assume that the mass and distance of the host star lens are $M_{\rm L} = 0.3\ M_{\odot}$ and $D_{\rm L} = 6\ \rm{kpc}$, and the distance of the source star is $D_{\rm S} = 8\ \rm{kpc}$.
Then, the angular and physical Einstein radii of the lens system are $\thetae = 0.32\ \rm{mas}$ and $r_{\rm E} = 1.9\ \rm AU$.
We adopt three different source stars including main-sequence, subgiant, and giant stars, which have radii of $1.0\ R_\odot$, $2.0\ R_\odot$, and $10.0\ R_\odot$, respectively.
The radii of the three source stars normalized to the Einstein radius are $\rho_{\star} = 0.0018, 0.0036,$ and $0.018$.
For high-magnification events, the effect of limb darkening of the finite source surface is not negligible, so we adopt a brightness profile for the source star of the form
\begin{equation}
{I(\theta)\over{I_0}} = {1 - \Gamma \left(1-{3\over{2}}{\rm cos}\theta \right) - \Lambda \left(1-{5\over{4}}{\rm cos}^{1/2}\theta \right) },
\end{equation}
where $\Gamma$ and $\Lambda$ are the linear and square-root coefficients and $\theta$ is the angle between the normal to the surface of the source star and the line of sight \citep{an02}.
We assume that the coefficients ($\Gamma$, $\Lambda$) of main-sequence, subgiant, and giant stars are (0.08, 0.52), (0.11, 0.51), and (0.21, 0.46), respectively.

Figure 1 shows the probability of occurrence of EWCP events with $\delta \leq 2\%$ as a function of the projected star-planet separation in units of $\thetae$, $s$, and planet/star mass ratio, $q$.
The physical separation, $d$, and planet mass in units of Earth mass, $m_{\rm p}$, are also presented in the figure, where they are determined by $d = r_{\rm E}s$ and $m_{\rm p} = qM_{\rm L}$, respectively.
In each panel, different shades of grey represent the areas with the probabilities of $\geq 10\%$, $\geq 40\%$, $\geq 80\%$, and $100\%$, respectively.
As one may expect, the probability increases as the mass ratio decreases and the separation decreases for $s < 1$ and increases for $s > 1$. 
From the figure, we find that for $\lesssim 100\ M_{\rm E}$ planets with separations of  $0.2\ {\rm AU} \lesssim d \lesssim 20\ {\rm AU}$, EWCP events with $\delta \leq 2\%$ in high-magnification events of  $A_{\rm max} \geq 100$ occur with a frequency of $> 50\%$.
This implies that high-magnification events of $A_{\rm max} \geq 100$ are mostly EWCP events of $\delta \leq 2\%$, thus it is important to resolve the EWCP events for the detection of many different planetary systems.
In Figure 1, the reason a bump occurs at $s = 1.0$ is that regions with small fractional deviations around the center of the resonant caustic are rather widely formed \citep{chung09}.
The vertical line of the $100\%$ probability in the figure represents the boundary of the lensing zone of $0.6 \lesssim s \lesssim 1.6$, where the planetary caustic is located within the Einstein ring.
Because of the planetary caustic, the probability of occurrence of $\delta \leq 2\%$ events cannot reach $100\%$ within the lensing zone, as shown in Figure 1.
However, if the finite source effect is strong, the probability can reach $100\%$ at $s \sim 1$, since there exist regions with small fractional deviations within the resonant caustic, as mentioned before (see the bottom panel of Figure 1).
Table 1 shows in detail the probability for the three source stars presented in Figure 1.

\subsection{\it Detectability }

To estimate the efficiency of detecting planets in EWCP events of $\delta \leq 2\%$, we compute the detectability defined as the ratio of the fractional deviation ($\delta$) to the photometric accuracy ($\sigma_{\rm ph}$), i.e.,
\begin{equation}
D = {|\delta|\over{\sigma_{\rm ph}}}, \ \ \  \sigma_{\rm ph} = {{\sqrt{AF_{\rm S} + F_{B}}}\over{(A - 1)F_{\rm S}}}, 
\end{equation}
where $F_{\rm S}$ and $F_{\rm B}$ represent the baseline flux of the lensed source star and the blended background flux, respectively.
We assume that the I-band absolute magnitudes of the main-sequence, subgiant, and giant stars are $M_I = 3.8, 3.0,$ and $0.0$.
We assume that the extinction toward the Galactic bulge is $A_I = 1.0$ and the blended flux $F_{\rm B}$ is equivalent to the flux of the background star with the apparent magnitude of $I = 20.0$.
The apparent magnitudes of the three source stars affected by the assumed extinction are $I = 19.3, 18.5,$ and $15.5$, respectively.
Considering typical Galactic bulge events, we also assume that the Einstein timescale is $\te = 20$ days. 

Based on the specification of KMTNet systems, we assume that the instrument can detect 31 photons $s^{-1}$ for a $I = 20.0$ star and the monitoring frequency is once per 10 minutes, having an exposure time of 2 minutes, and the lower limit of the photometric accuracy is $0.001$,  which corresponds to the value at $I = 13.8$ mag (Atwood et al. 2012).
We assume that the planetary signal is detectable if $D \geq 3$.
We also assume that the planet is detected only if the planetary signal with $D \geq 3$ is detected at least five times during the event.
The five points with $D \geq 3$ do not need to be consecutive, because events with this requirement have $\Delta \chi^{2} \gtrsim 350$, which satisfies the detection threshold for high-magnification events \citep{gould10}.
We note that since the single lensing magnification $A_0$ is unknown in actual observations, it is determined from the best-fit single lens model to the observed planetary lensing event.
The lensing parameters of the best-fit single lens event are obtained by a $\chi^2$ minimization method \citep{chung06}.
Figure 2 shows the planet detection efficiency of EWCP events with $\delta \leq 2\%$ for three different source stars, as a function of $s$ and $q$. 
From the figure, we find the following results.
\begin{enumerate}
\item[1.]
EWCP events of $q \gtrsim 10^{-5}$ generally have two separations with the maximum detection efficiency due to the $s \leftrightarrow 1/s$ degeneracy, and the efficiency decreases as $s$ becomes smaller and/or larger than the maximum efficiency separation that changes with the planet mass.
This is because the photometric accuracy of the observation systems is limited, thus it does not allow the efficiency for each $q$ to continuously increase as $s$ decreases for $s < 1$ and increases for $s > 1$.
This result implies that the planet detection in EWCP events occurs only within a limited separation range which depends on the planet mass.
\item[2.]
For main-sequence and subgiant stars, $\gtrsim 1\ M_{\rm E}$ planets in EWCP events of $\delta \leq 2\%$ can be detected $> 50\%$ in a certain range that changes with the planet mass.
The range for the two stars is presented in Tables 2 and 3.
\end{enumerate} 

EWCP events are caused by close, intermediate, and wide planetary systems with low-mass planets and close and wide planetary systems with massive planets.
These planetary systems are quite different from those already detected, which are mostly intermediate planetary systems with $s \sim 1$ regardless of the planet mass.
Therefore, the above results imply that a much greater variety of planetary systems can be significantly detected in the near future. 
We also compare the estimated detection efficiency with the probability in the same range.
As a result, the efficiency for $\gtrsim 1\ M_{\rm E}$ planets increases to $\sim 70\%$, $\sim 80\%$, and $\sim 80\%$ for main-sequence, subgiant, and giant stars, respectively.
This means that $\gtrsim 1\ M_{\rm E}$ planets located within the certain range can almost be detected by KMTNet.
The efficiency compared with the probability, i.e., the ratio of the efficiency to the probability, is presented in Tables 2, 3, and 4. 
In Figure 2, the white solid and dashed lines represent the set of points where the probabilities of occurrence of events with $\delta \leq 1\%$ and $\delta \leq 0.5\%$ are both $80\%$.
From the figure, we find that KMTNet can readily resolve up to EWCP events of $\delta = 0.5\%$.

The planet detection efficiency is sensitive to the Einstein timescale of a lensing event.
This is because under the condition of the limited monitoring frequency of a system, the decrease of the timescale gives rise to a decrease of the chance of detecting planets during the event.
We thus test the change of the detection efficiency depending on the Einstein timescale.
Figure 3 represents the detection efficiency changing with the Einstein timescale for the planetary system of $s = 0.5$ and $q = 10^{-4}$.
The efficiency dramatically increases until $\te < 10$ days, but it becomes constant for $\te \gtrsim 10$ days, as shown in the figure.
This means that the estimated efficiency for three source stars is valid up to EWCP events with $\te \sim 10$ days, whereas for events with $\te < 10$ days it considerably decreases with $\te$.

\section{DISCUSSION}

Most high-magnification events of $A_{\rm max} \geq 100$ are EWCP events of $\delta \leq 2\%$ and KMTNet is capable of resolving the EWCP events, as shown in the results of Figures 1 and 2.
However, the KMTNet is more favorable for the detection of events caused by the planetary caustic than those caused by the central caustic (high-magnification events).
This is because the KMTNet plans to do 24-hour continuous observation with a constant exposure time of 2 minutes and the exposure time is applied for stars of $13 < I < 20$ mag \citep{kim10}.
Hence, if a source star is highly magnified by $I < 13$ mag, it is easy for the KMTNet to be saturated around the peak of the high-magnification event with a 2-minute exposure time.
This means that it is difficult to detect planets in EWCP events of bright stars like giant stars using KMTNet.
The estimated detection efficiency for main-sequence and subgiant stars also includes those cases where the source stars are highly magnified by $I < 13$ mag, and thus the efficiency would decrease in real observations.
However, since one considers only high-magnification events with $A_{\rm max} \geq 100$, KMTNet has a chance of detecting EWCP events of more dark stars of $I \gtrsim 20$ mag.
The result of the test for the $I \gtrsim 20$ stars shows that the KMTNet can resolve up to EWCP events of $I \sim 22.0$ stars.
Figure 4 shows an example light curve of a $I = 21.9$ star highly magnified by the planetary lens system of $s = 2.3$ and $q = 2.4\times 10^{-4}$.
The planetary lensing event has a planetary signal of $\delta \lesssim 2\%$ and the planetary signal can be detected by KMTNet, according to the assumed detection conditions.
Moreover, if follow-up spectroscopic observations would be carried out while observing high-magnification events using KMTNet, the chemical information for many faint bulge stars could be obtained, and it would be very helpful in studying the origin of the Galactic bulge.

\citet{yee13} mentioned that based on detected planetary lensing events, the detection threshold for central caustic events seems higher than for planetary caustic events.
To confirm whether this is true or not, the detection of many more events caused by the central and planetary caustics is needed.
In particular, the detection of more planetary caustic events is needed because only six of 25 microlensing planets have been detected in the planetary caustic events (Beaulieu et al. 2006; Sumi et al. 2010; Muraki et al. 2011; Bennett et al. 2012; Poleski et al. 2013; Furusawa et al. 2013; Tsapras et al. 2013 ). 
Fortunately, a vast number of planetary caustic events will be detected by the observation strategy of KMTNet.
In addition, Figure 2 shows a total range of central caustic events of $A_{\rm max} \geq 100$ that can be detected by KMTNet.
The region includes both EWCP events (grey region) and non-EWCP events (white region marked as $``A"$), which represents events of $\delta > 2\%$.
KMTNet can readily detect events within the white region, because it is much easier to detect events of $\delta > 2\%$ than those of $\delta \leq 2\%$.
A very wide range of grey and white regions implies that a large number of central caustic events (i.e., high-magnification events) will also be detected by KMTNet.
Therefore, by using the KMTNet, one can find out whether the detection threshold for high-magnification events is higher than that for planetary caustic events and determine more accurate detection thresholds of both events.

\section{CONCLUSION}

We have estimated the probability of occurring EWCP events of $\delta \leq 2\%$ in high-magnification events of $A_{\rm max} \geq 100$ and the efficiency of detecting planets in the EWCP events using the next-generation ground-based observation project, KMTNet.
From this study, we found that the EWCP events occur with a frequency of $> 50\%$ in the case of $\lesssim 100\ M_{\rm E}$ planets with separations of $0.2\ {\rm AU} \lesssim d \lesssim 20\ {\rm AU}$.
This implies that most of high-magnification events of $A_{\rm max} \geq 100$ are EWCP events of $\delta \leq 2\%$, and thus it is important to resolve the EWCP events for the detection of many different planetary systems.
We found that for main-sequence and subgiant stars, $\gtrsim 1\ M_{\rm E}$ planets in EWCP events of $\delta \leq 2\%$ can be detected $> 50\%$ in a certain range that varies depending on the planet mass.
However, it is difficult to detect planets in EWCP events of bright stars like giant stars, because it is easy for KMTNet to be saturated around the peak of the EWCP events with a constant exposure time.
EWCP events are caused by close, intermediate, and wide planetary systems with low-mass planets and close and wide planetary systems with massive planets.
Therefore, we expect that a much greater variety of planetary systems than those already detected, which are mostly intermediate planetary systems with $s \sim 1$ regardless of the planet mass, will be significantly detected in the near future.

\begin{figure}[t]
\epsscale{1.0}
\plotone{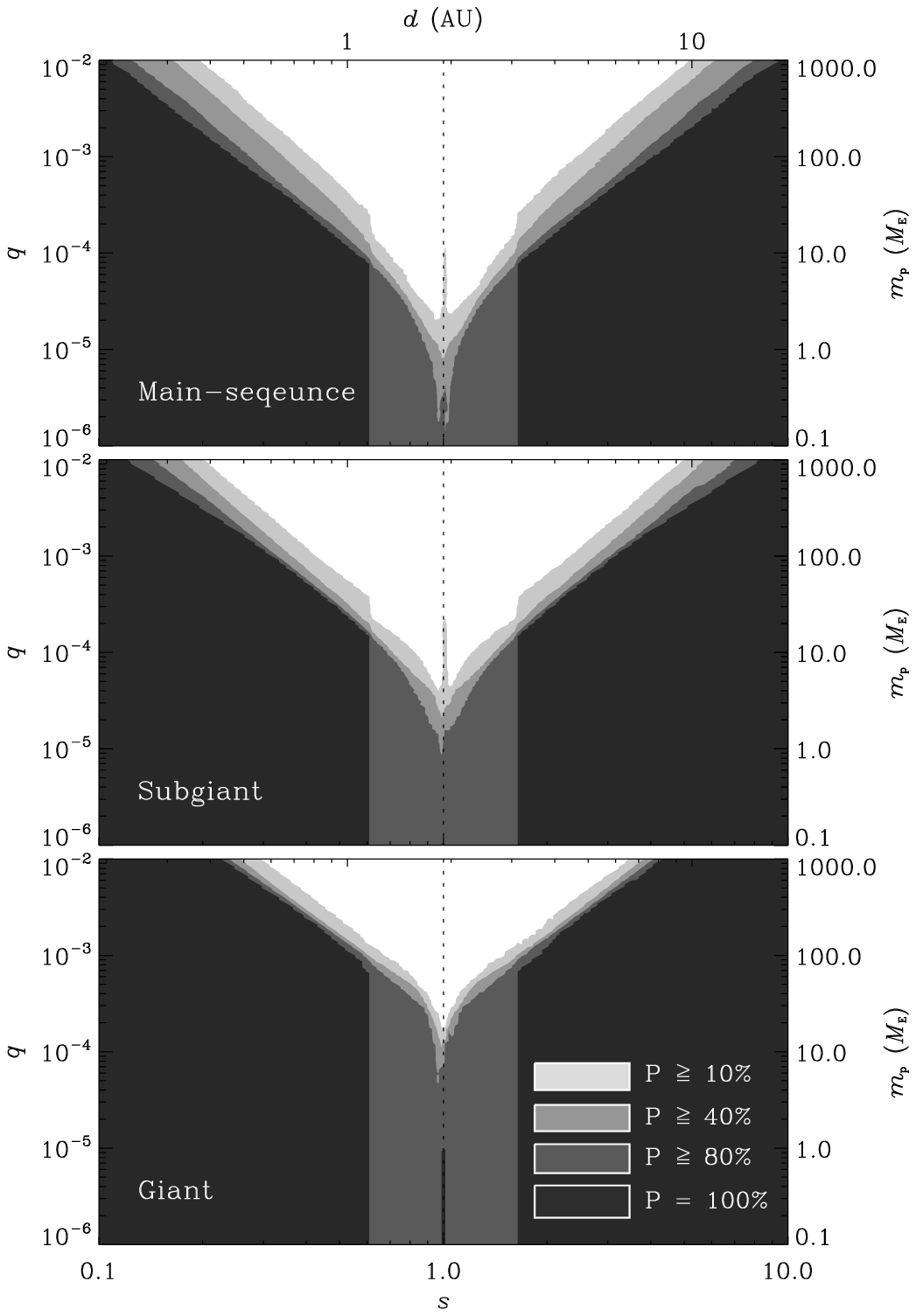}
\caption{\label{fig:one}
Probability of occurrence of EWCP events with $\delta \leq 2\%$ in high-magnification events of $A_{\rm max} \geq 100$ for three different source stars, main-sequence, subgiant, and giant stars, as a function of the projected star-planet separation in units of $\thetae$, $s$, and plant/star mass ratio, $q$.
The physical separation, $d$, and planet mass in units of Earth mass, $m_{\rm p}$, are also presented.
The radii of the three source stars in units of $\thetae$ are $\rho_{\star} = 0.0018, 0.0036,$ and $0.018$, respectively.
In each panel, different shades of grey represent the areas with the probabilities of $\geq 10\%$, $\geq 40\%$, $\geq 80\%$, and $100\%$, respectively.
The vertical dot line indicates the separation of $s = 1$.
}
\end{figure}

\begin{figure}[t]
\epsscale{1.0}
\plotone{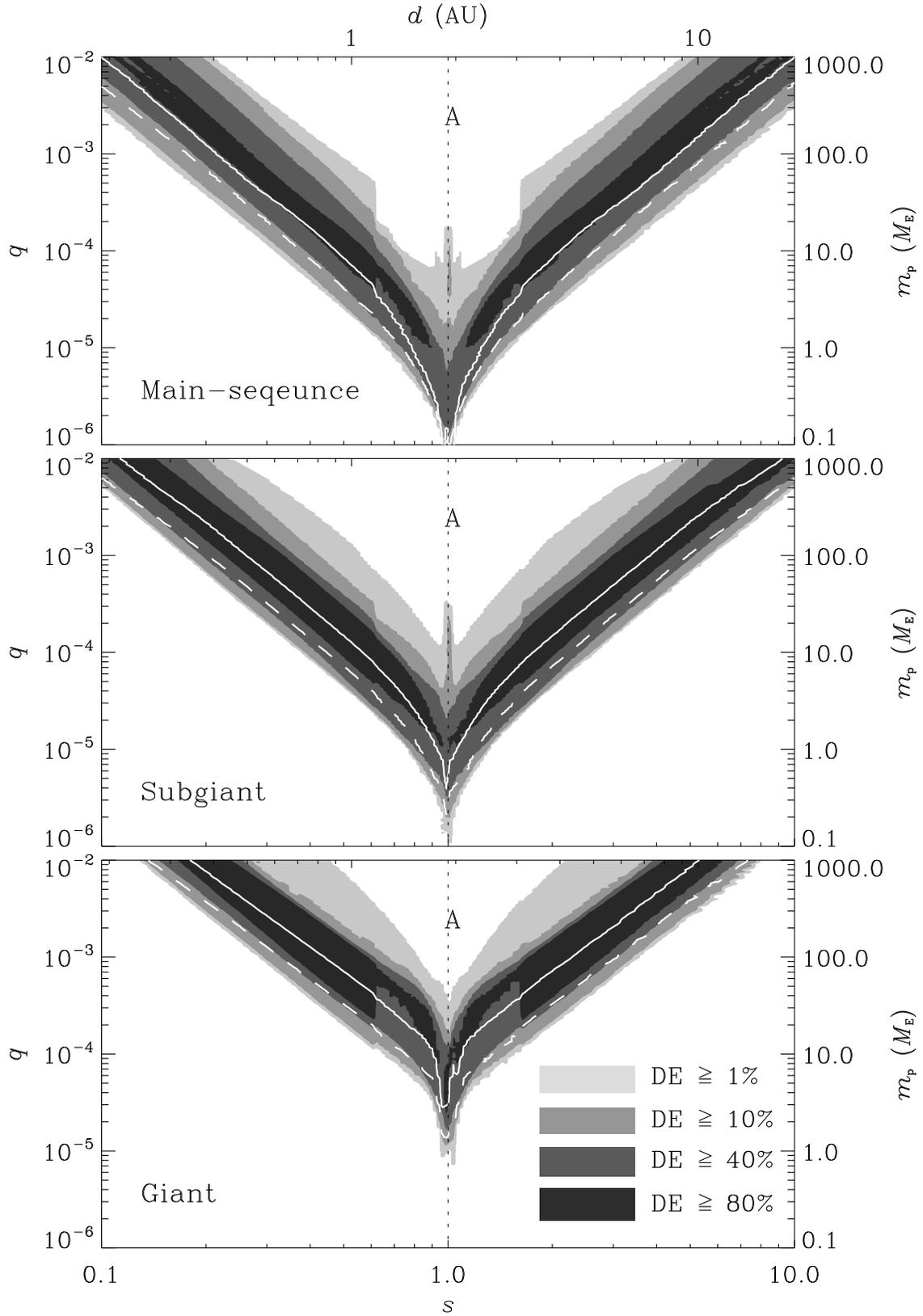}
\caption{\label{fig:three}
Planet detection efficiency of EWCP events with $\delta \leq 2\%$ for three different source stars.
Different shades of grey represent the areas with the efficiencies of $\geq 1\%, \geq 10\%, \geq  40\%,$ and $\geq 80\%$, respectively.
The white solid and dashed lines represent the set of points where the probabilities of occurring events with $\delta \leq 1\%$ and $\delta \leq 0.5\%$ are both $80\%$.
The white region marked as $``A"$ represents the region in which high-magnification events with $\delta > 2\%$ (not EWCP events) occur.
}
\end{figure}

\begin{figure}[t]
\epsscale{1.0}
\plotone{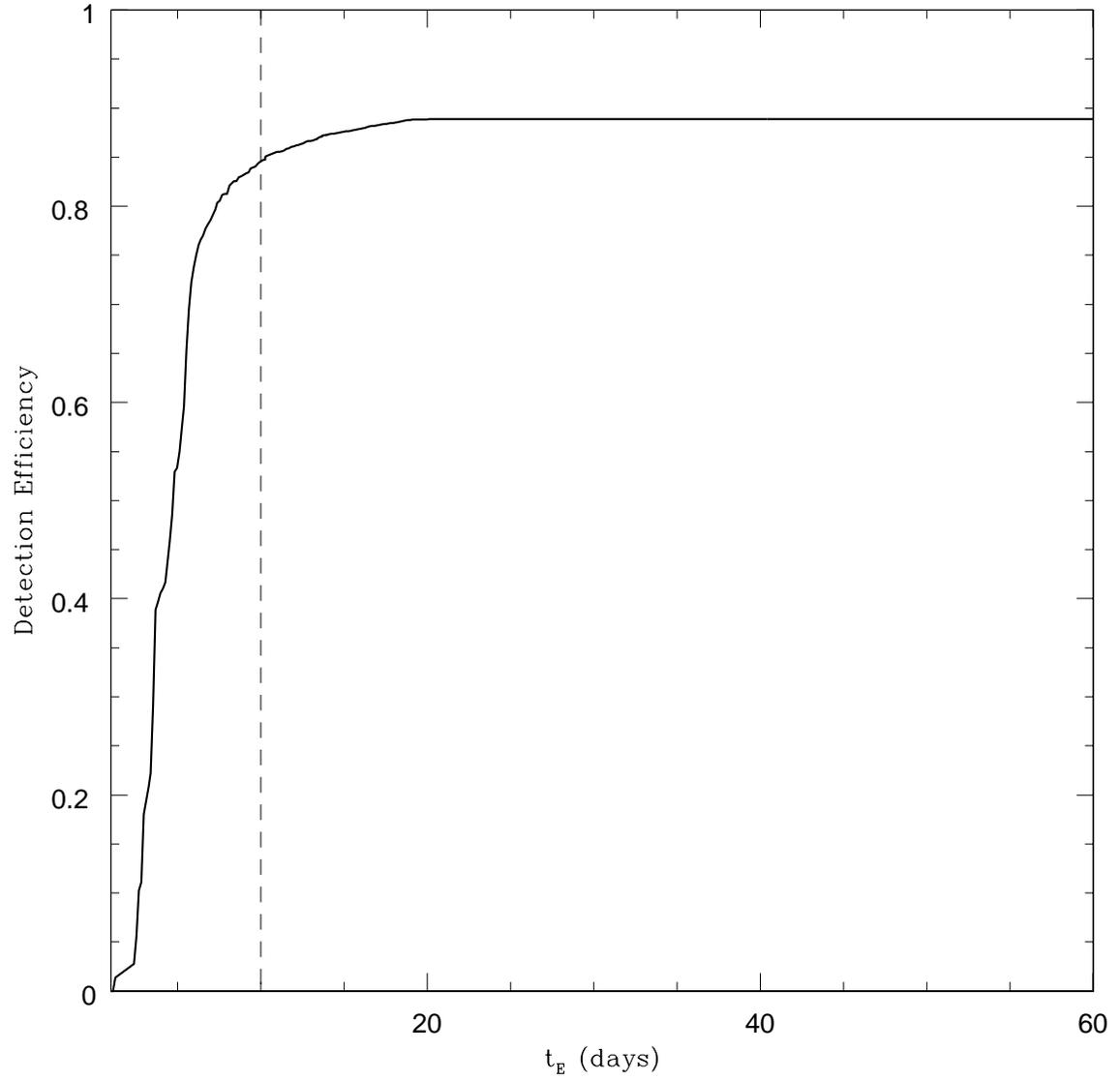}
\caption{\label{fig:four}
Detection efficiency as a function of the Einstein timescale for the planetary system of $s = 0.5$ and $q = 10^{-4}$.
The vertical dashed line indicates a timescale of $\te = 10$ days.
}
\end{figure}

\begin{figure}[t]
\epsscale{1.0}
\plotone{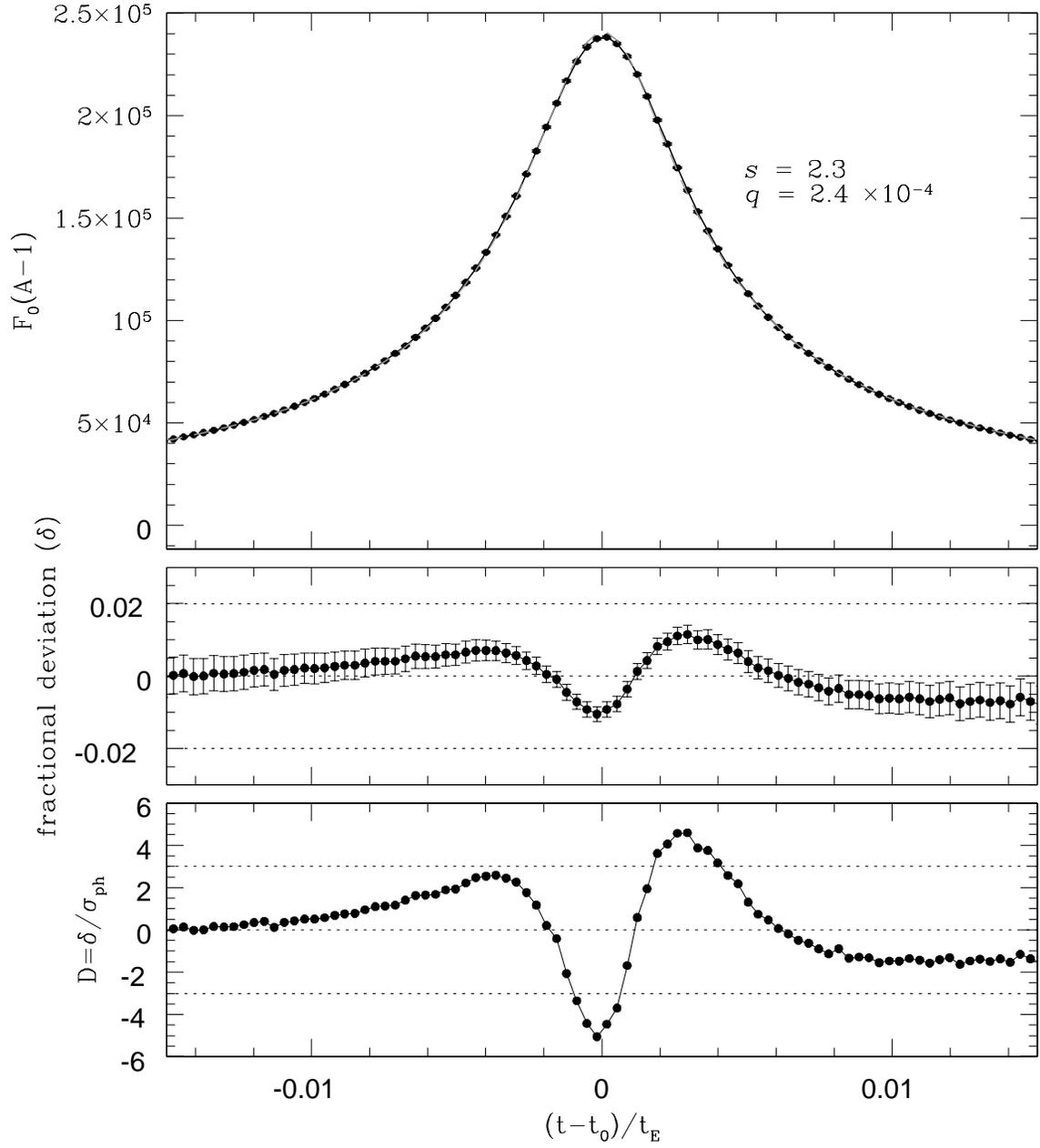}
\caption{\label{fig:four}
Example light curve of a $I = 21.9$ star highly magnified by the planetary lens system of $s = 2.3$ and $q = 2.4\times 10^{-4}$.
The black solid and grey dashed lines in the top panel are the light curves of the planetary lensing and best-fit single lensing events.
}
\end{figure}

\begin{deluxetable}{lccccccc}
\tablecaption{Probability of the occurrence of EWCP events.\label{tbl-one}}
\tablewidth{0pt}
\tablehead{
& \multicolumn{7}{c}{Probability ($\%$)} \\
\cline{3-7}\\
Planet mass && main-sequence && subgiant && giant
}
\startdata
300.0$M_{\rm E}$ && 35.9 && 40.1  && 61.3 \\
100.0$M_{\rm E}$ && 51.6 && 56.2  && 78.9 \\
10.0$M_{\rm E}$   && 79.7 && 85.7  && 96.2 \\
5.0$M_{\rm E}$     && 85.7 && 91.5  && 96.5 \\
1.0$M_{\rm E}$     && 95.3 && 96.9  && 97.0 \\
0.5$M_{\rm E}$     && 97.1 && 97.5  && 97.2
\enddata
\end{deluxetable}

\begin{deluxetable}{lccccccccccccc}
\tablecaption{Detection efficiency for a main-sequence star \label{tbl-two}}
\tablewidth{0pt}
\tablehead{
Planet mass && range (AU) &&& Detection Efficiency (DE) && Probability (P) && Ratio (DE/P) \\
&& ($s < 1$) &&&\\ 
&& ($s > 1$) &&&  ($\%$) && ($\%$) && ($\%$)
}
\startdata

300.0$M_{\rm E}$ && $0.2 \lesssim d \lesssim 0.6 $     &&& 53.9 &&  75.3  && 71.6 \\
                             && $6.6 \lesssim d \lesssim 17.8 $   &&& 54.4 &&  75.4  && 72.1 \\\\ 
100.0$M_{\rm E}$ && $0.3 \lesssim d \lesssim 0.8 $     &&& 54.7 &&  76.7  && 71.3 \\
                            && $4.6 \lesssim d \lesssim 12.4 $   &&& 54.7  &&  75.8  && 72.2 \\\\ 
10.0$M_{\rm E}$   && $0.6 \lesssim d \lesssim 1.6 $     &&& 50.9 &&  68.7  && 74.1 \\
                             && $2.3 \lesssim d \lesssim 5.7 $     &&& 50.5 &&  68.5  && 73.7 \\\\ 
5.0$M_{\rm E}$     && $0.8 \lesssim d \lesssim 1.7 $     &&& 51.7 &&  70.0  && 73.9 \\
                             && $2.1 \lesssim d \lesssim 4.6 $     &&& 51.3 &&  69.5  && 73.8 \\\\ 
1.0$M_{\rm E}$     && $1.3 \lesssim d \lesssim 2.8 $     &&& 58.0 &&  75.7  && 76.6 \\\\ 
0.5$M_{\rm E}$     && $1.5 \lesssim d \lesssim 2.4 $     &&& 57.4 &&  85.0  && 67.5 
\enddata
\tablecomments{
The physical Einstein radius of the assumed lens system is $r_{\rm E} = 1.9\ \rm AU$.
The range represents the region with the efficiency $\gtrsim 10 \%$, where it is rather wide, as shown in Figure 2.
The range is divided into $s < 1$ and $s > 1$ only for the cases where the region with the efficiency $\gtrsim 10\%$ is clearly separated based on $s = 1$.
}
\end{deluxetable}

\begin{deluxetable}{lccccccccccccc}
\tablecaption{Detection efficiency for a subgiant star \label{tbl-three}}
\tablewidth{0pt}
\tablehead{
Planet mass && range (AU) &&& Detection Efficiency (DE) && Probability (P) && Ratio (DE/P) \\
&& ($s < 1$) &&& \\ 
&& ($s > 1$) &&& ($\%$) && ($\%$) && ($\%$)
}
\startdata
300.0$M_{\rm E}$ && $0.2 \lesssim d \lesssim 0.6 $     &&& 64.3 &&  79.7  && 80.7 \\
                             && $6.4 \lesssim d \lesssim 15.5 $   &&& 64.6 &&  80.1  && 80.6 \\\\ 
100.0$M_{\rm E}$ && $0.3 \lesssim d \lesssim 0.8 $     &&& 65.8 &&  80.9  && 81.3 \\
                             && $4.5 \lesssim d \lesssim 10.7 $   &&& 66.0 &&  80.5  && 82.0 \\\\
10.0$M_{\rm E}$   && $0.7 \lesssim d \lesssim 1.6 $     &&& 64.8 &&  81.2  && 79.8 \\
                             && $2.3 \lesssim d \lesssim 4.9 $     &&& 64.9 &&  81.7  && 79.4 \\\\
5.0$M_{\rm E}$     && $0.9 \lesssim d \lesssim 1.8 $     &&& 62.6 &&  79.1  && 79.1\\
                             && $2.0 \lesssim d \lesssim 3.9 $     &&& 62.4 &&  79.0  && 79.0\\\\
1.0$M_{\rm E}$     && $1.5 \lesssim d \lesssim 2.5 $     &&& 55.4 &&  85.6  && 64.7\\\\
0.5$M_{\rm E}$     && $1.7 \lesssim d \lesssim 2.1 $     &&& 39.0 &&  90.1  && 43.3 
\enddata
\end{deluxetable}

\begin{deluxetable}{lccccccccccccc}
\tablecaption{Detection efficiency for a giant star \label{tbl-four}}
\tablewidth{0pt}
\tablehead{
Planet mass && range (AU) &&& Detection Efficiency (DE) && Probability (P) && Ratio (DE/P) \\
&& ($s < 1$) &&& \\ 
&& ($s > 1$) &&&  ($\%$) && ($\%$) && ($\%$)
}
\startdata
300.0$M_{\rm E}$ && $0.4 \lesssim d \lesssim 0.8 $     &&& 68.2 &&  87.8  && 77.7 \\
                             && $4.4 \lesssim d \lesssim 9.8 $     &&& 67.0 &&  89.6  && 74.8 \\\\
100.0$M_{\rm E}$ && $0.6 \lesssim d \lesssim 1.3 $     &&& 69.9 &&  86.7  && 80.6 \\
                             && $2.9 \lesssim d \lesssim 6.5 $     &&& 71.0 &&  89.1  && 79.7 \\\\
10.0$M_{\rm E}$   && $1.3 \lesssim d \lesssim 2.8 $     &&& 57.6 &&  81.7  && 70.5 \\\\
5.0$M_{\rm E}$     && $1.6 \lesssim d \lesssim 2.2 $     &&& 66.9 &&  83.5  && 80.1 

\enddata
\end{deluxetable}

\acknowledgments
Numerical simulations were performed by using a high performance computing cluster at the KASI (Korea Astronomy and Space Science Institute).
This work was supported by the KASI grant 2014-1-400-06.

\end{document}